\documentclass[aps,prb,twocolumn,groupedaddress]{revtex4-1}

\usepackage{graphicx}
\usepackage[space]{grffile}
\usepackage{dcolumn}
\usepackage{bm}
\usepackage{subfigure}
\usepackage{mathrsfs}
\usepackage{booktabs}
\usepackage{notoccite}
\usepackage{float}
\usepackage{placeins}
\usepackage{enumerate}
\usepackage{gensymb}
\usepackage{amsmath}
\usepackage{color}
\usepackage{siunitx}

\newcommand{\rom}[1]{\uppercase\expandafter{\romannumeral #1\relax}}

\begin{document}

\title{Quantum--continuum calculation of the surface states \\ and electrical response of silicon in solution}

\author{Quinn Campbell}
\email{quinn.campbell@psu.edu}
\author{Ismaila Dabo}
\affiliation{Department of Materials Science and Engineering, Materials Research Institute, and Penn State Institutes of Energy and the Environment, The Pennsylvania State University, University Park, PA 16802, USA}

\begin{abstract}
A wide range of electrochemical reactions of practical importance occur at the interface between a semiconductor and an electrolyte.  We present an embedded density-functional theory method using the recently released self-consistent continuum solvation (SCCS) approach to study these interfaces. In this model, a quantum description of the surface is incorporated into a continuum representation of the bending of the bands within the electrode. The model is applied to understand the electrical response of ten relevant surface terminations for silicon electrodes in solution, providing microscopic insights into the low-voltage region, where surface states determine the electrification of the semiconductor electrode.
\end{abstract}

\maketitle

\section{\label{sec:introduction}Introduction}

Predicting the electrical response and stability of semiconductor--solution interfaces is of central relevance to a wide array of electrochemical and photoelectrochemical systems.  These interfaces are involved in the photocatatytic splitting of water,\cite{Kudo2009} the photoreduction of carbon dioxide into hydrocarbons,\cite{Roy2010} the electrochemical etching of semiconductor surfaces,\cite{Tenne1983,Jin2004,Peng2006} the storage of energy at metal oxide electrodes,\cite{Xia2012} and the use of quantum dots as biological markers.\cite{Gill2008}  The pivotal role of semiconductor electrodes at the frontier between solid state physics and electrochemistry provides a compelling motivation to study their behavior in solution.

Density-functional theory has been used to search for new photocatalysts,\cite{Bhatt2015,Castelli2012a,Castelli2015} assess the alignment of the valence and conduction bands of semiconductor electrodes with the redox potentials of species in solution,\cite{Ping2015a,Ping2015,Cheng2010,Wu2011,Yang2012} determine reaction pathways for photoelectrochemical reactions,\cite{Setvin2016,Li2016,Cheng2016,Shen2010} and elucidate the dynamical interactions of the solvent molecules with the surface of the semiconductor.\cite{Oshikiri2006,Rivest2014,Lee2014,Tritsaris2014, Kharche2014,Cheng2014} The same calculations can be applied to predict the electrical response of semiconductor--solution interfaces as long as they account for the long-range decay of the electrostatic potential within the semiconductor depletion region.  However, electrostatic screening in doped semiconductors is much less effective than in metals,\cite{Rajeshwar2007} causing the interfacial electric field to penetrate up to 10--10$^3$ nm into the electrode for typical dopant concentrations of $10^{16}$--$10^{18}$ cm$^{-3}$. These length scales render the first-principles simulation of the interface computationally demanding.

\begin{figure}[h!]
	\includegraphics[width=\columnwidth]{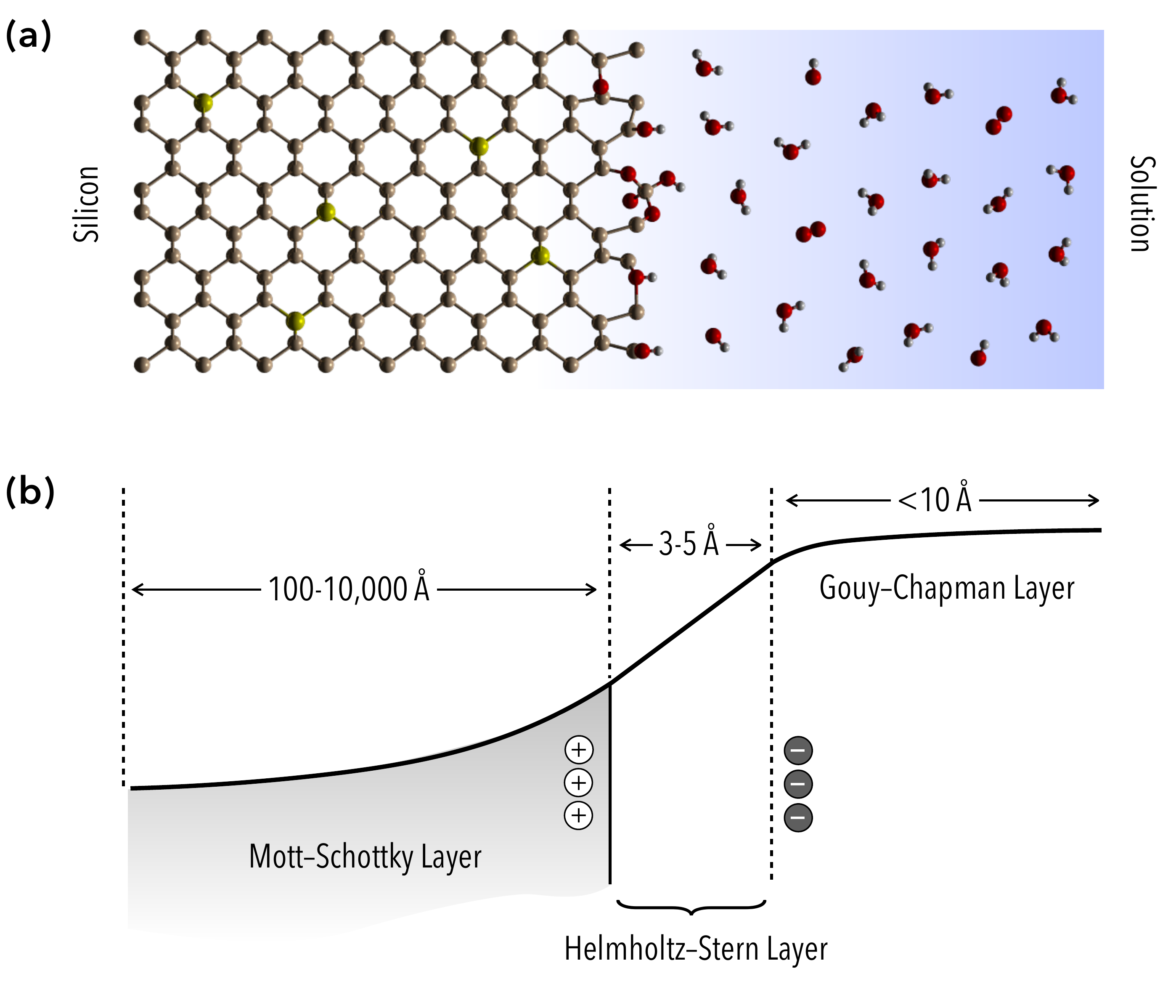}
	\caption{\label{fig_1} \small (a) Atomic-level view of a semiconductor--solution interface. (b) Electrostatic profile across the semiconductor--solution interface, showing the band bending of the electronic bands on the semiconductor side, described by an extended Mott--Schottky layer, and the electrical double layer on the solution side, represented by a Helmholtz--Stern layer in series with a Gouy--Chapman layer of oppositely charged ions.}
\end{figure}

Therefore, it is necessary to develop efficient models that will capture the essential features of a semiconductor--electrolyte interface at reduced computational cost. To this end, we exploit and further develop the self--consistent continuum solvation (SCCS) approach proposed by Andreussi et al.\cite{Andreussi2012} to simulate semiconductor electrodes under applied voltage in electrolytic media. While this method has been successful in modeling metal electrodes,\cite{Bonnet2013, Sementa2016a, Montemore2016, Keilbart2017, Weitzner2017} no previous study has focused on describing band bending at semiconductor electrodes using a self--consistent continuum solvation approach.

To illustrate band bending, the electrostatic profile of a semiconductor--electrolyte interface is shown in Fig.~\ref{fig_1}.  In order to reach equilibrium, the electrochemical potential of the semiconductor and the solution need to be equal. Since the excess charge is accommodated within the semiconductor by a low concentration of dopants, the electrostatic potential is seen to decay gradually across the extended depletion region (the Mott--Schottky layer). This ideal picture is made more complicated, however, by the presence of surface states, which result from the adsorption of ionic species.  By trapping charge at the interface, these surface states lead to much stronger electrostatic screening within the semiconductor.  It is the goal of this work to develop a quantum--continuum model for  simulating the response of semiconductor--electrolyte interfaces under electrification, including the influence of the surface states.  We apply the model to prototypical silicon electrodes in solution in an effort to elucidate the connection between their electrical response and surface structure at the molecular level. 

\section{Method}

\subsection{Interface energy}

\begin{figure}
	\centering \includegraphics[width=\columnwidth]{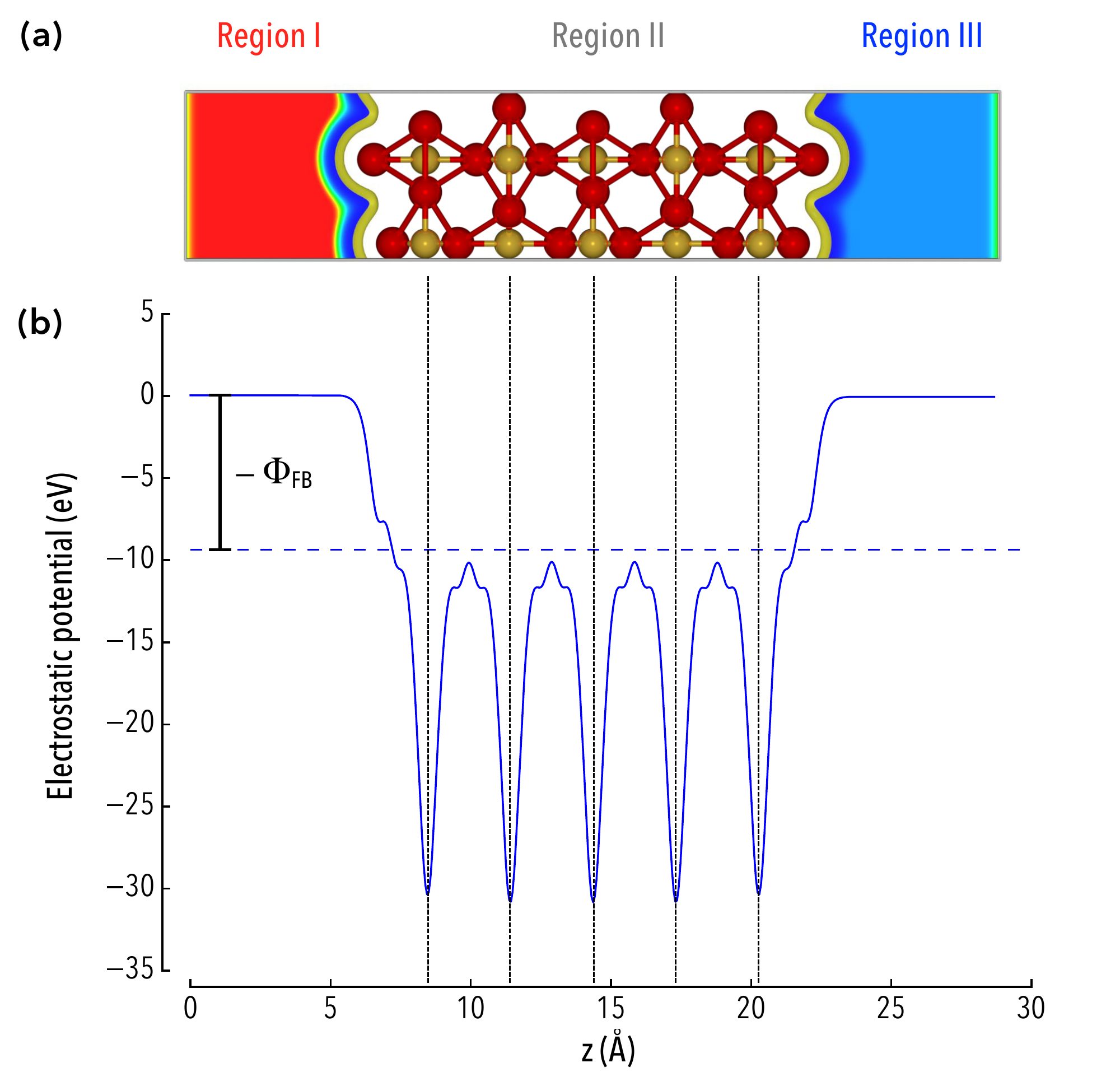}
	\caption{\label{fig_2}\small (a) Partition of a rutile ${\rm SiO}_{2}$(110)--electrolyte interface into three regions. Region I represents the continuum bulk semiconductor section, Region II the quantum surface of the semiconductor, and Region III the continuum electrolyte solution. Both Regions I and III extend infinitely. The colors indicate the changing dielectric constant in the simulation; red corresponds a dielectric constant of $\sim$3.9 for ${\rm SiO}_{2}$ and blue corresponds to a dielectric constant of $\sim$78 for water at room temperature. (b) Profile of the electrostatic potential across the ${\rm SiO}_{2}$--electrolyte interface with the dotted horizontal line representing the Fermi level $\epsilon_{\rm F}$ of the slab.  Having set the potential to zero at the boundaries of the cell, the flatband potential $\Phi_{\rm FB}$ equals the negative of the Fermi level.}
\end{figure}

The first step in constructing the model is to partition the system into the three regions shown in Fig.~\ref{fig_2}. Region I represents the bulk of the semiconductor, which will be modeled at the continuum level. Region II corresponds to the surface of the electrode; this region will be described quantum mechanically to represent the adsorbed species and resulting surface states. Region III denotes the electrolytic solution, which consists of a diffuse distribution of ions in a polarizable continuum.  Also illustrated is the flatband potential $\Phi_{\rm FB}$, which corresponds to the difference between the asymptotic value $\varphi_0$ of the potential inside Region I and the Fermi energy of the neutral slab:
\begin{equation}
\Phi_{\rm FB} = (\varphi_0 - \epsilon_{\rm F})/e.
\end{equation}

Having defined the three regions, the free energy of the system is written as
\begin{equation}
F = F_{\rm I} + F_{\rm II} + F_{\rm III} - \dfrac{1}{2 } \int d\bm{r} \epsilon_0\epsilon(\bm{r}) \left|\nabla\varphi(\bm{r}) \right|^{2},
\end{equation}
where $\epsilon$ denotes the space-dependent dielectric permittivity across the interface, and $F_{\rm I}$, $F_{\rm II}$, and $F_{\rm III}$ stand for the free energies of the continuum space charge, quantum surface slab, and continuum electrolyte, respectively.

This dielectric permittivity is defined using the self-consistent continuum solvation model (SCCS).\cite{Andreussi2012} In this model, a dielectric cavity is created at each lateral facet.  On the semiconductor side, the local dielectric permittivity can be written as $\epsilon (\bm{r}) =\exp[(\zeta(\bm{r}) -\sin(2\pi\zeta(\bm{r}))/2\pi)\ln\epsilon_{\rm I}]$ where $\epsilon_{\rm I}$ is the dielectric constant of the bulk of the semiconductor and $\zeta(\bm{r}) =(\ln \rho_{\rm max} - \ln\rho(\bm{r}))/(\ln \rho_{\rm max} -\ln \rho_{\rm min}) $ is used as a smooth switching function, marking the transition between the quantum and continuum regions. Here, $\rho_{\rm min}$ and $\rho_{\rm max}$ serve as the density thresholds specifying the inner and outer isocontours of the dielectric cavity. We employ the same parametrization in Region III, replacing $\epsilon_{\rm I}$ with $\epsilon_{\rm III}$, the dielectric constant of the electrolyte.  We specifically use  $\rho_{\rm max} = \num{5e-3}$ a.u. and $\rho_{\rm min} = 10^{-4}$ a.u. for our calculations.

Focusing first on Region I, the contribution to the free energy can be expressed in terms of the local density of negative charge carriers $n$ and positive charge carriers $p$ as the sum of electrostatic and entropic terms:
\begin{eqnarray}
F_{\rm I} &=&\int d\bm{r} (\varphi(\bm{r})+\epsilon_{\rm V}-\epsilon_{\rm F})p(\bm{r})  - T s(p(\bm{r}),p_{\rm d}(\bm{r})) \nonumber \\
&-&\int d\bm{r} (\varphi(\bm{r})+\epsilon_{\rm C}-\epsilon_{\rm F})n(\bm{r}) + T s(n(\bm{r}),n_{\rm d}(\bm{r})) \nonumber\\
&+&\int d\bm{r} \varphi(\bm{r})(n_{\rm d}(\bm{r})-p_{\rm d}(\bm{r}))
\label{eq:free_energy_I}
\end{eqnarray}
with
$$
\textstyle s(f,f_{\rm d}) = - k_{\rm B} \left[ \left(f_{\rm d} - f  \right) \ln\left(1 - \dfrac{f}{f_{\rm d}}\right) + f  \ln\left(\dfrac{f}{f_{\rm d}}\right)\right].
$$
In Eq.~\eqref{eq:free_energy_I}, $\epsilon_{\rm F}$ denotes the Fermi energy, $\epsilon_{\rm V}$ is the electronic energy at the top of the valence band, and $\epsilon_{\rm C}$ is the  energy at the bottom of the conduction band.  Furthermore, it is assumed that the donor and acceptor levels are shallow so that their energy levels sit at the band edges.  Moreover, $s$ stands for the Fermi--Dirac entropy; it depends locally on the smooth switching functions
\begin{eqnarray}
n_{\rm d}(\bm{r}) = \dfrac{\mathscr N}{2} \left[{\rm erfc}\left(\dfrac{z-z_{\rm I}}{\sigma_{\rm I}}\right) +1\right]\nonumber \\
p_{\rm d}(\bm{r}) = \dfrac{\mathscr P}{2} \left[ {\rm erfc}\left(\dfrac{z-z_{\rm I}}{\sigma_{\rm I}}\right)+1\right]
\end{eqnarray}
with ${\mathscr N}$ and ${\mathscr P}$ being the concentrations of electron-donating and electron-accepting defects, and $z_{\rm I }$ and $\sigma_{\rm I}$ being the location and spatial extent of the transition between the semiconductor and surface.

Likewise, the free energy of the electrolyte can be expressed in terms of the concentrations $c_+$ and $c_-$ of the positive and negative ions as
\begin{eqnarray}
F_{\rm III} &= & \int d\bm{r} \varphi(\bm{r})c_+(\bm{r}) - T \sigma(c_+(\bm{r}),c^\circ(\bm{r})) \nonumber \\
&-&\int d\bm{r} \varphi(\bm{r})c_-(\bm{r}) + T \sigma(c_-(\bm{r}),c^\circ(\bm{r}))
\label{eq:free_energy_III}
\end{eqnarray}
with
$$
\sigma(c_{\pm},c^\circ) = - k_{\rm B}  \left[  f \ln\left(\dfrac{c_{\pm}}{c^\circ}\right) - c_{\pm} \right]
$$
In Eq.~\eqref{eq:free_energy_III}, we take a symmetric 1:1 ionic solution, in which the maximal ion concentration is defined as
\begin{equation}
c^\circ(\bm{r}) = \dfrac{\mathscr C}{2}\left[{\rm erfc}\left(\dfrac{z_{\rm III}-z}{\sigma_{\rm III}}\right)+1\right],
\end{equation}
where $\mathscr C$ is the equilibrium ionic concentration inside the electrolyte, and $z_{\rm III}$ and $\sigma_{\rm III}$ are the location and spread of the frontier between the surface and electrolyte.

Finally, the free energy $F_{\rm II}$ is expressed as a Kohn--Sham functional of the density of the electrons $\rho_-$ and distribution of the atomic cores $\rho_+$:
\begin{eqnarray}
F_{\rm II}  &= &  T_{\rm s} + E_{\rm Hxc} - \theta {\mathscr S} -\int d\bm{r} \varphi \rho_+ -(\varphi-\epsilon_{\rm F})\rho_-,
 \end{eqnarray}
where $T_{\rm s}$ is the kinetic energy of the auxiliary system within the independent-electron mapping and $E_{\rm Hxc}$ is the sum of the Hartree and exchange-correlation energies.  The  electronic temperature and the entropy of the electronic smearing are denoted $\theta$ and ${\mathscr S}$, respectively.

\subsection{Interface electrostatics}

With the expression of the free energy in hand, the equilibrium charge density can be obtained by variations with respect to the occupations of the doping levels, ionic concentrations, and electrostatic potential, yielding the following electrostatic problem:
\begin{eqnarray}
	\label{eq:full-Poisson}
	\nabla \left( \epsilon_0 \epsilon({\bm r}) \nabla \varphi(\bm{r}) \right) &=& p({\bm r})  -n({\bm r}) -p_{\rm d}({\bm r})  +n_{\rm d}({\bm r})  \\ & +& c_+({\bm r})   - c_-({\bm r}) + \rho_+({\bm r}) - \rho_-({\bm r})\nonumber,
\end{eqnarray}
where the source terms can be expressed as
\begin{equation}
n(\bm{r}) = n_{\rm d}(\bm{r})\left[ 1 + \exp\left(\dfrac{\varphi(\bm{r})+\epsilon_{\rm C}-\epsilon_{\rm F}}{k_{\rm B}T}\right) \right]^{-1}
\end{equation}
\begin{equation}
p(\bm{r}) = p_{\rm d}(\bm{r})\left[ 1 + \exp\left(\dfrac{\epsilon_{\rm F}-\epsilon_{\rm V}-\varphi(\bm{r})}{k_{\rm B}T}\right) \right]^{-1}
\end{equation} 
\begin{equation}
c_{\pm}(\bm{r}) = c^\circ(\bm{r}) \exp\left(\mp \dfrac{\varphi({\bm r})}{k_{\rm B}T}\right)
\end{equation}
Here, it is understood that $\rho_-$ is obtained by solving the self-consistent Kohn-Sham equation for a given distribution $\rho_+$ of the atomic cores.

For a $n$-type semiconductor, using the Boltzman distribution, these equations become
\begin{equation}
p({\bm r}) = p_{\rm d}({\bm r}) = 0
\end{equation}
\begin{equation}
n({\bm r}) = n_{\rm d}({\bm r})\exp\left( \dfrac{\varphi_0-\varphi({\bm r})}{k_{\rm B}T}\right),
\end{equation}
where $\varphi_0$ stands for the asymptotic value of the potential in Region I.  Conversely, for a $p$-type semiconductor, we can write
\begin{equation}
n({\bm r}) = n_{\rm d}({\bm r}) = 0
\end{equation}
\begin{equation}
p({\bm r}) = p_{\rm d}({\bm r})\exp\left(\dfrac{\varphi({\bm r})-\varphi_0}{k_{\rm B}T}\right).
\end{equation}

Furthermore, it is important to note that deep inside the semiconductor region, the electrostatic potential obeys the one-dimensional Poisson equations:
\begin{equation}
\dfrac{d^2 \varphi}{dz^2} = \dfrac{\mathscr N}{\epsilon_0 \epsilon_{\rm I}}\left[1-\exp\left(\dfrac{\varphi_0-\varphi}{k_{\rm B}T}\right)\right]
\end{equation}
\begin{equation}
\dfrac{d^2 \varphi}{dz^2} = \dfrac{\mathscr P}{\epsilon_0 \epsilon_{\rm I}}\left[\exp\left(\dfrac{\varphi-\varphi_0}{k_{\rm B}T}\right)-1\right]
\end{equation}
under conditions of $n$-type and $p$-type doping, respectively.  In the long-range limit where $\varphi$ approaches $\varphi_0$, these equations imply that
\begin{equation}
\left(\dfrac{d \varphi}{dz}\right)^2 = \dfrac{2\mathscr N}{\epsilon_0 \epsilon_{\rm I}} \left[ \varphi -  \varphi_0 + k_{\rm B}T \left( e^{\tfrac{\varphi_0-\varphi}{k_{\rm B}T}} -1 \right) \right]
 \label{eq:Mott-Schottky-1}
\end{equation}
\begin{equation}
\left(\dfrac{d \varphi}{dz}\right)^2 = \dfrac{2 \mathscr P}{\epsilon_0 \epsilon_{\rm I}} \left[ \varphi_0 -  \varphi + k_{\rm B}T \left( e^{\tfrac{\varphi-\varphi_0}{k_{\rm B}T}} -1 \right) \right].  \label{eq:Mott-Schottky-2}
\end{equation}
As explained in Sec.~\ref{sec:band-bending}, these expressions are of central utility in describing the bending of the electronic bands and overcome the length scales that characterize electrostatic screening in the depletion region of the electrode.

\subsection{Band bending}

\label{sec:band-bending}

Equation \eqref{eq:full-Poisson} can be solved by implementing a fully self-consistent solution of the electrostatic problem.  We plan to implement this method in the continuation of this study.  For the moment, we use a simpler implementation to assess the model.  The details of this approach are presented below.

To obtain the equilibrium charge-voltage distribution of the system, we start by specifying a total charge for the electrode, from which the potential of the system can be found. To this end, we place a plane of charge $q_{\rm I}$ in Region I, representing the defect charge in the bulk of the semiconductor, and another plane of countercharge $q_{\rm III}$ in Region III, representing the ionic charge of the electrolyte.  Accordingly, an explicit charge $q_{\rm II}$ is added to the slab of Region II to fulfill charge neutrality:
$$
q_{\rm I}+q_{\rm II}+q_{\rm III}=0.
$$

This planar setup provides an accurate approximation of the electrolytic side (Region III) as long as the ionic concentrations ($10^{20}$--$10^{21}$ cm$^{-3}$) in the electrolyte are significantly larger than typical doping concentrations in the semiconductor ($10^{16}$--$10^{18}$ cm$^{-3}$).  This means that most of the potential drop takes place in the bulk of the semiconductor (Region I), making a plane of countercharge a reliable representation of the response of the electrolyte (Region III) (the Stern model). 

Furthermore, on the semiconductor side, adding a plane of countercharge within Region I does not lead to any loss of generality in the solution of the problem within Region II and Region III.  This can be seen by noting that once the Helmholtz plane of countercharges in Region III is set and the asymptotic boundary conditions of Poisson's equation inside the electrolyte is fixed, the Fermi energy $\epsilon_{\rm F}$ and charge density $\rho_-$ at the surface are fully determined; they do not depend on the specific shape of the defect charge profile in virtue of Gauss' law.  In other words, the charge distribution and potential profile on the right hand side provide an accurate description of the interaction of the electrode with solution.  

\begin{figure}
	\centering \includegraphics[width=\columnwidth]{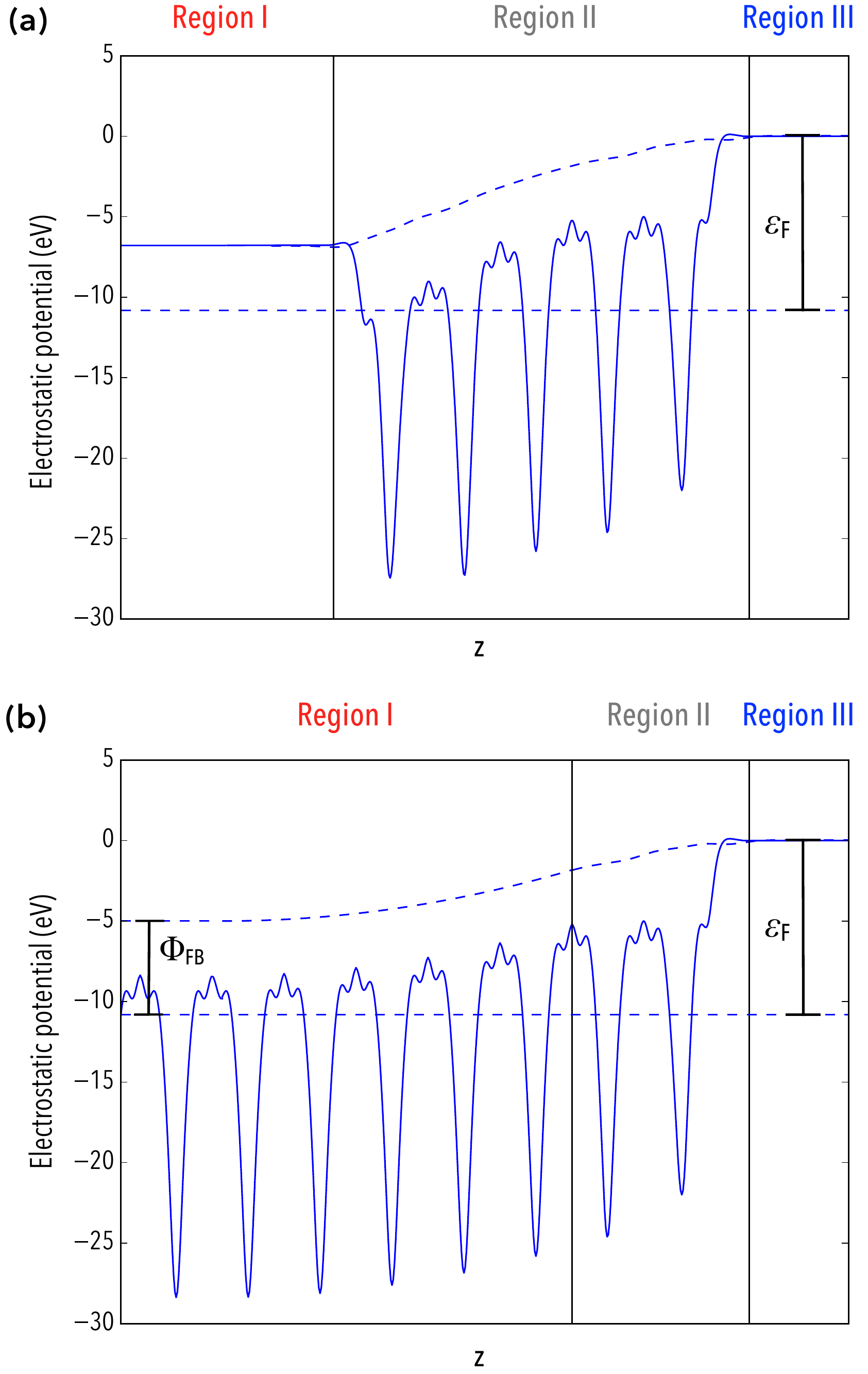}
	\caption{\label{fig_3}\small (a) The potential of a charged slab with planes of countercharge on each side, creating a potential drop. The dotted line represents the electrostatic potential $\bar \varphi$ of the charged slab subtracted from that of a slab with zero charge as shown in Fig.~\ref{fig_2}. (b) A cutoff value $z_{\rm c}$ corresponding to the inflection of the potential $\bar \varphi$ is determined. To the left of this cutoff a Mott--Schottky extrapolation is applied, as shown by the new dotted line. By examining several different charge distributions, the specific distribution where the Fermi levels match is found. The width of the depletion region is shortened here for illustrative purposes and would normally extend for several nanometers.}
\end{figure}

Although the electrostatic profile in Regions II and III is accurate, the potential in Region I is still not a reliable representation of the electrostatics of a semiconductor electrode where Region I consists of an infinite extension of Region II.  To accurately describe the electrostatic potential in Region I, a cutoff plane is introduced within the slab as illustrated in Fig.~\ref{fig_3}, and the electrostatic potential to the left of the cutoff is set to follow the solution of the electrostatic equations of a continuum dielectric. The region to the right of the cutoff (but still within the explicit slab) will be taken as Region II, representing the surface states. The position of the cutoff plane is a user defined value which should correspond to around the inflection of the potential $\bar \varphi$. It should be far enough within the slab that a bulk like state has occurred, making a smooth transition from the surface states to the bulk of the semiconductor. From the value of the electrostatic potential and its derivative at the cutoff plane, the Fermi level of the bulk of the semiconductor in Region I can be easily determined from Eqs.~\eqref{eq:Mott-Schottky-1} and \eqref{eq:Mott-Schottky-2}:
\begin{equation}
	\varepsilon_{\rm F,I} = \bar \varphi_0 - e\Phi_{\rm FB}
\end{equation}
with
\begin{equation}
	\bar \varphi_0 = \bar {\varphi}(z_{\rm c}) -  k_{\rm B}T - \dfrac{\epsilon_0 \epsilon_{\rm I} }{2 \mathscr N} \left(\dfrac{d \bar \varphi}{dz}(z_{\rm c})\right)^2
\end{equation} 
\begin{equation}
	\bar \varphi_0 = \bar {\varphi}(z_{\rm c}) +  k_{\rm B}T + \dfrac{\epsilon_0 \epsilon_{\rm I} }{2 \mathscr P} \left(\dfrac{d \bar \varphi}{dz}(z_{\rm c})\right)^2
\end{equation} 
for $n$-type and $p$-type semiconductor, respectively.  In these equations, the cutoff position $z_{\rm c}$ represents the location of the frontier between Region I and Region II in Fig.~\ref{fig_3}b and $\bar {\varphi}$ is the difference between the electrostatic potential of the charged slab and that of the neutral slab, corresponding to the dashed line in Fig.~\ref{fig_3}a. The bulk potential of the electrode can then be taken as $\varepsilon_{\rm F,I}$.

Finally, to find the equilibrium state of the charge electrode, we impose that the Fermi level of the bulk of the semiconductor must equal the Fermi level of the quantum slab:
\begin{equation}
	\varepsilon_{\rm F,I} = \varepsilon_{\rm F,II} .
	\label{eq:Fermi_balance}
\end{equation}
By satisfying this condition, the charge density of the electrode can be calculated as a function of voltage, and the surface state density can finally be obtained as the total charge on the right hand side of the frontier defined by $z_{\rm c}$ divided by the elementary charge. Different algorithms can be used to find the conditions of matching Fermi levels such as a dichotomy procedure. The procedure we use is defined in Sec.~\ref{sec:computational-details}. 

This protocol enables us to determine how surface states and adsorption affect the potential profile and capacitance of the system. This approach is, however, limited in a few respects. First, the position of the frontier between Region I and Region II defined by $z_{\rm c}$ may affect the asymptotic value $\bar \varphi_0$ of the potential $\bar {\varphi}$ describing the overall trend of the potential $\varphi $ across the interface.  This variation is, however, small and can be easily evaluated from $\Delta  \bar \varphi_0  = \frac{\Delta z_{\rm c}}{{\mathscr L}_{\rm I}}\frac{d \bar \varphi}{dz}(z_{\rm c}) $ with ${\mathscr L}_{\rm I}$ being the electrostatic screening length of the semiconductor.  Since ${\mathscr L}_{\rm I}$ is on the order of 10--10$^3$ nm, the sensitivity of $\bar \varphi_0$ to $z_{\rm c}$ is negligible under relevant doping conditions.  Second, the range of charge that can be tested is dependent on the size of the slab. In fact, if the voltage drop between the two countercharge planes is larger than the band gap of the material, unphysical charge transfer by Zener tunneling will take place between the two sides of the slab. In these cases, smaller slabs serve to reduce the voltage drop for a constant Helmholtz charge density. It is important, however, to verify that the Fermi level converges with respect to the size of the slab used for the calculation. Third, the solution is essentially planar within Region I and III. This assumes that both the bulk of the semiconductor and the solution see no variation in the planar directions. For a few applications such as quantum dots, this assumption may need to be revisited. Nevertheless, this is a valid assumption within most applications of interest to first-principles surface electrochemistry. 

\section{Computational details}

\label{sec:computational-details}

\begin{figure}
	\includegraphics[width=\columnwidth]{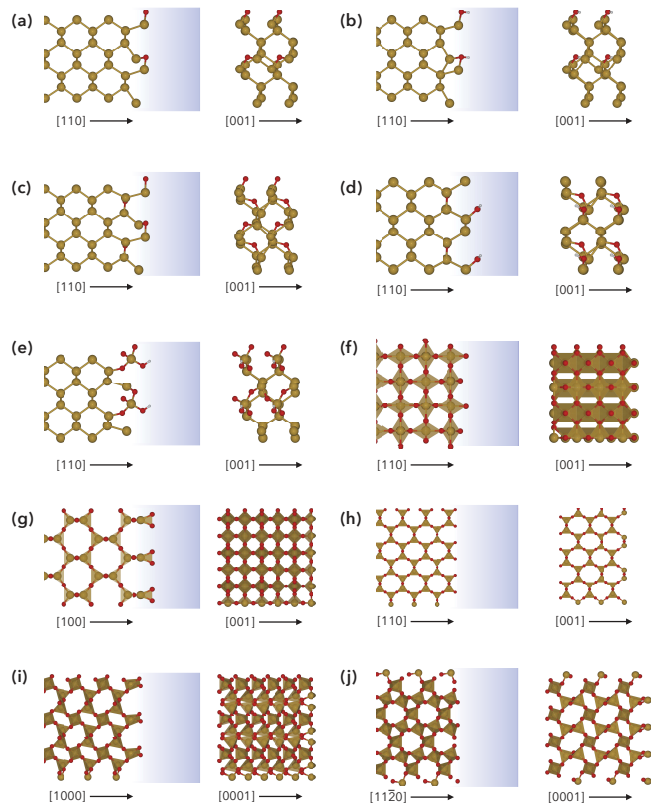}
	\caption{\label{fig_4} \small Lateral and top views of representative surface terminations for silicon: (a) Si(110) with oxygen (O*), (b) Si(110) with a hydroxyl group (O* + H*), (c) Si(110) with two oxygens absorbed into the surface layers (2O*), (d) Si(110) with an oxygen absorbed into the second layer with and adsorbed hydroxyl group (2O* + H*), (e) Si(110) with a ${\rm SiO}_{4}$ tetrahedron terminated by a hydrogen (4O* + H*), (f) Rutile ${\rm SiO}_{2}$(110), (g) Cristobalite ${\rm SiO}_{2}$(100), (h) Cristobalite ${\rm SiO}_{2}$(110), (i) Quartz ${\rm SiO}_{2}$(1000), and (j) Quartz ${\rm SiO}_{2}$(11$\bar 2$0).}
\end{figure}

Density-functional theory calculations are performed using the {\sc pw} code of the Quantum-Espresso distribution.\cite{Giannozzi2009}  As shown in Fig. \ref{fig_4}, surface slabs of 1 $\times$ 1 Si(110) and rutile, cristobalite, and quartz ${\rm SiO}_{2}$ (110) and cristobalite and quartz ${\rm SiO}_{2}$ (100) are constructed with a slab width of 5 layers, which is sufficient to give converged Fermi levels within 0.01 eV. The slab is centered in the supercell with a vacuum height of 7 \AA \ to ensure convergence of the atomic forces within a few meV/\AA.   We use ultrasoft pseudopotentials with the Perdew--Burke--Ernzerhof  parameterization of exchange--correlation interactions.\cite{Perdew1996} The cutoffs of kinetic energy of the charge density and electrons are set at 50 Ry and 750 Ry, respectively. The Brillioun zone is sampled with a shifted 5 $\times$ 5 $\times$ 1 Monkhorst--Pack grid and 0.03 Ry of Marzari--Vanderbilt smearing.\cite{Marzari1997} 

As explained above, the electrostatic response of the electrolyte interface is modeled using the {\sc environ} module with the parameterizations developed for water.\cite{Andreussi2012} Several  surface configurations of the Si and ${\rm SiO}_{2}$ surfaces are examined. In finding the equilibrium structure for these configurations, the three layers closest to Region I are frozen to create a bulk-like condition. The final relaxed positions are then used in the semiconductor--interface model discussed above. To find the equilibrium charge distribution between the surface states and the bulk of the semiconductor, several partitions of the charge $q_{\rm III}$ between $q_{\rm I}$ and $q_{\rm II}$ are tested for a fixed $q_{\rm III}$. In explicit terms, for each total electrode charge, 11 different partitions were considered: one where 0\% of the electrode charge is in Region I and 100\% in Region II, one where 10\% of the electrode charge is Region I and 90\% of the electrode charge is in Region II, and so on. Using a fixed cutoff position, $z_{\rm c}$, we find the charge distribution that minimizes the difference in Fermi level between Regions I and II from Eq.~\eqref{eq:Fermi_balance}.  A dopant concentration of $10^{18}$ cm$^{-3}$ was used along with a dielectric constant of silicon as $ \epsilon_{\rm I} = 11.7 $ and a dielectric constant of water of $ \epsilon_{\rm III} = 78.3$. The results of these simulations are presented and discussed in Sec.~\ref{sec:results-discussion}.

\section{Results and Discussion}

\label{sec:results-discussion}

\begin{figure}
	\includegraphics[width=\columnwidth]{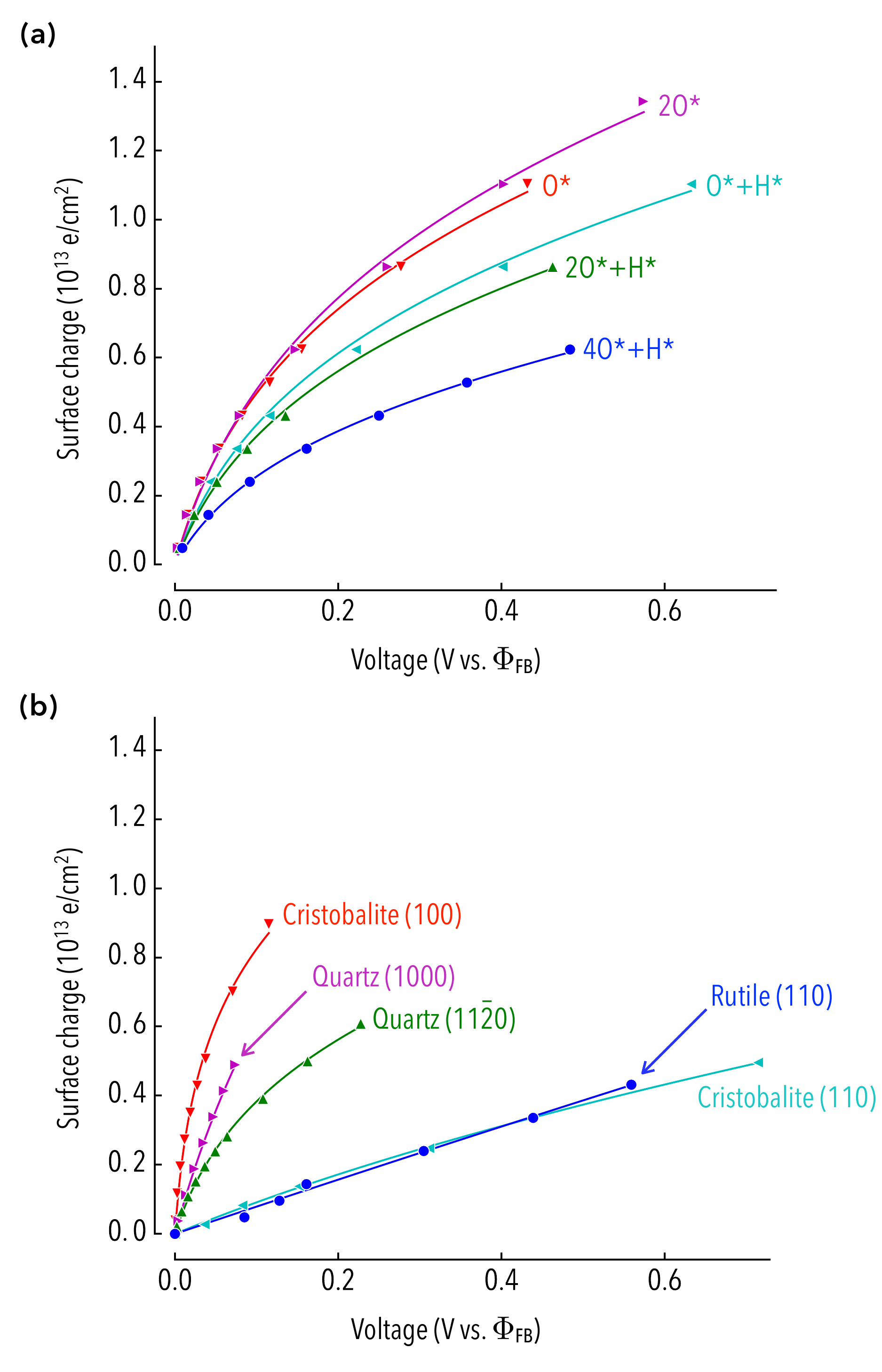}
	\caption{\label{fig_5} \small (a) The total charge versus voltage curves for Si (110) structures. (b) The total charge versus potential curves for ${\rm SiO}_{2}$ structures. The lines correspond to the fitted trends of an empirical model that consists of an ideal Mott-Schottky semiconductor
	in series with a linear capacitor representing the surface states.}
\end{figure}

Silicon electrodes represent an important test for the proposed method.  Experiment shows that an amorphous layer of ${\rm SiO}_{2}$ forms on the surface of Si in contact with water.\cite{Runyan2013,Graf1989} However, the underlying driving force for this oxidation is not clearly understood and can vary with both pH and voltage.\cite{Leung2009,Li2004} In response, several different terminations of silicon were tested. The results of these calculations can be seen in Fig.~\ref{fig_5}. 

When charge is placed on a solvated silicon slab with planes of countercharge for charge neutrality, all of the electronic charge accumulates at the semiconductor edge.\cite{SupMat} This contrasts  with the expected electron distribution throughout a semiconductor electrode with surface states, implying that all the charge on a silicon semiconductor slab would accumulate deep inside the bulk of the semiconductor. As a result, there is no combination of charges $q_{\rm I}$ and $q_{\rm II}$ that equalizes the Fermi levels in Region I and Region II [Eq.~\eqref{eq:Fermi_balance}]. This shows that no equilibrium charge distribution exists between the surface of silicon and its bulk state, offering insight into the instability of pure silicon in water at low potentials.\cite{Nikolaychuk2014} From this, we conclude that no surface states are likely to form on a pure silicon surface. Thus, some significant contribution from adsorption is expected for the electrification of surface states on a silicon electrode in water.

To gain insight into the role that surface states play on the charge-voltage curves semiconductor--solution interfaces, we use a simple model that consists of describing the charge--voltage behavior of a semiconductor electrode as an ideal Mott--Shottky semiconductor in series with a metal surface state. This gives an overall capacitance of the form
$1/{\mathscr C}_{\rm I+II}(\Phi) =  1/{\mathscr C}_{\rm I}(\Phi_{\rm B}) + 1/{\mathscr C}_{\rm II}$, where ${\mathscr C}_{\rm I}$ is the capacitance of Region I obtained from Eq.~\eqref{eq:Mott-Schottky-1}, ${\mathscr C}_{\rm II}$ is the capacitance of the interface region, and $\Phi_{\rm B}$ is the potential drop across Region I (the Schottky barrier).  In this model, ${\mathscr C}_{\rm II}$ does not depend on the potential $\Phi$, whereas ${\mathscr C}_{\rm I}$ depends on it through $\Phi_{\rm B}$.  Furthermore, we describe the relation between $\Phi_{\rm B}$ and $\Phi$ in the vicinity of the flatband potential, that is, for a small amount of charge accumulation at the electrode, as $\Phi_{\rm B} =  \gamma_{\rm I} (\Phi-\Phi_{\rm FB})$, where  $\gamma_{\rm I}$ is the fraction of the total potential drop that occurs within the semiconductor, limited to values between 0 and 1.  The value of  $\gamma_{\rm I}$ accounts for the contribution from surface states to the overall electrical response; the value of  $\gamma_{\rm I}$ decreases with increasing charge buildup in the surface states.  The parameters of the fitted curves, shown in Fig.~\ref{fig_5}, are reported in Table \ref{Table_1}.

\begin{table}
	\small
	\label{Table_1}%
	\centering
		\begin{tabular*}{0.8\columnwidth}{l @{\extracolsep{\fill}} ll}		     \\
			& $\gamma_{\rm I}$ (\%)& ${\mathscr C}_{\rm II}$ ($\mu$F cm$^{-2}$) \\ 
			\hline \\
		    Si+O*  & 3.75 & 78.9  \\
		    Si+O*+H* & 5.39 & 54.1  \\
		    Si+2O* & 3.38 &  75.2  \\
		    Si+2O*+H & 5.92 & 38.9 \\
		    Si+4O*+H* & 11.7 & 25.3  \\
		    Rutile (110) & 0.03 & 1.28 \\
		    Cristobalite (100) & 1.69 & 786 \\
		    Cristobalite (110) & 8.23 & 1.89  \\
		    Quartz (1000) & 1.95 & 41.3  \\
		    Quartz (11$\bar{2}$0) & 4.71 & 30.1  \\ \\ \hline
               \end{tabular*}%
               
        	\caption{Fitted surface state properties for the 10 surface configurations tested. The percentage of the total potential drop that takes place within Region I, the bulk of the semiconductor, is represented by $\gamma_{\rm I}$ (\%). The capacitance of the surface state, assuming a metal like distribution, is represented as ${\mathscr C}_{\rm II}$.}
	\end{table}

Simulations using different surface terminations of silicon show a strong buildup of charge upon increasing the potential, as can be seen in Fig.~\ref{fig_5}a. This reflects the fact that nearly all the charge is trapped in Region II of the material, as confirmed by the observation that the fraction $\gamma_{\rm I}$ of the potential drop taking place within the semiconductor is low for all of these structures. In general, adding hydroxyl groups to the surface causes the charge to be more distributed within the material than with just an oxygen added. This is supported by noting that the fraction of the potential drop within the semiconductor is higher for structures with hydrogen added than the corresponding structures without hydrogen. One likely explanation for this is that negatively charged adsorbates lead to a trapping of positive charge near the surface. Conversely, adding hydrogen to the surface reduces the electronegativity of the adsorbate and allows for more long-ranged charge distribution inside the semiconductor. This is further shown with the Si+4O*+H* structure, which has the most charge distribution within the electrode. Under typical conditions, however, the type of adsorbate at the silicon surface seems to have a moderate effect on the overall trend; in all these curves, a large initial buildup of charge due to surface states is followed by a much slower buildup dominated by the bulk of the semiconductor. All of the silicon adsorbate materials have a capacitance ${\mathscr C}_{\rm II}$ on the order of 10s of $\mu$F/cm$^2$.

In contrast, simulations with ${\rm SiO}_{2}$ terminations show large differences in the resulting charge--voltage response. In general, these structures have a much more distributed charge profile. This is reflected in Fig.~\ref{fig_5}b by the lowered charge density in comparison to the silicon adsorbate structures. In particular, the rutile ${\rm SiO}_{2}$(110) structure and cristobalite ${\rm SiO}_{2}$(110) structure present a stark difference with the other shown structures. This can be attributed to surface states with a much lower capacitance. This leads to much shallower growth of the charge--voltage curve. It should be noted that the value of ${\mathscr C}_{\rm II}$ for Cristabolite (110) is unphysically high; this is not due to the quantum continuum model, but rather indicative of the limitations of a two parameter fitting function. Another important aspect to note is the change in electrode behavior is a function of the exposed surface. Changing from the (100) to (110) orientation for a cristobalite and quartz ${\rm SiO}_{2}$ structure leads to a lower accumulation of charge. This further underscores the importance of the exposed facet in determining the voltage-dependent charge distribution across the interface. In a traditional Mott--Schottky model, the specific surface termination would not change the charge--voltage response. 

It should be noted that for surfaces that quickly grow positive with the application of a small amount of potential, it is expected that negatively charged species from the solution would adsorb at the surface. This would result in surface oxidation until a passivating oxide layer forms, providing insights into the experimentally observed formation of an oxide layer when silicon is in contact with water. \cite{Runyan2013,Graf1989}  For a more complete comparison with experiment, it would be necessary to perform simulations on large-scale amorphous surface terminations under applied voltage and controlled pH. This will be the subject of a study in the continuation of this work. 

The results presented here differ  from the ideal Mott--Schottky picture by providing a detailed description of charge accumulation at low potential where the surface states dominate the electrochemical properties of the electrode. 

\section{Conclusion}

Semiconductor--electrolyte interfaces encompass numerous applications at the frontier of solid state physics and electrochemistry. We have presented a method to embed first-principles calculations of surface states between a Mott--Schottky description of band bending within the semiconductor and the Helmholtz representation of the surrounding electrolyte. We have applied the method on different surface terminations for silicon with a focus on Si and ${\rm SiO}_{2}$ structures. These simulations provide a comprehensive atom-level understanding of the experimentally observed electrification of silicon electrodes in water, suggesting the rapid accumulation of positive charge at the surface of solvated silicon electrodes leading to the formation of an oxide layer that shifts the potential of charge neutrality to more positive voltages and ultimately prevents further oxidation. This method is ideally positioned to examine the low-potential regime where surface state charges dominate the electrification of the electrode in a manner not captured by the Mott--Schottky theory alone.  Future work will focus on the implementation and distribution of algorithms to determine the three-dimensional charge distribution between the bulk of the semiconductor, the surface states, and the electrolyte for predicting the structure and response of semiconductor--electrolyte interfaces under electrochemical conditions. 

\begin{acknowledgements}
The authors acknowledge primary support from the National Science Foundation under Grant DMR-1654625, and partial support from the 3M Graduate Fellowship and Penn State University Graduate Fellowship.
\end{acknowledgements}

\end{document}